\documentclass{appolb}
\usepackage[utf8]{inputenc}
\usepackage{bm}
\usepackage{bbold}
\usepackage{mathtools}
\usepackage[toc,page]{appendix}
\usepackage{dsfont}
\usepackage{combelow}
\usepackage[colorlinks=true,linkcolor=blue,citecolor=blue, urlcolor=blue]{hyperref} 
\usepackage{tikz}
\usepackage{graphicx}
\usepackage{subfigure}
\usepackage{scalerel,amssymb}
 
\def\del{\partial}

\def\sst{\scriptscriptstyle}

\long\def\comment#1{ }

\def\0{{\boldsymbol 0}}

\def\Kc{{\cal K}}

\def\pt{p_{\sst T}}
\def\pT{p_{\sst T}}

\newcommand{\beq}{\begin{eqnarray}}
\newcommand{\eeq}{\end{eqnarray}}
\newcommand{\be}{\begin{eqnarray*}}
\newcommand{\ee}{\end{eqnarray*}}
\newcommand{\rmd}{{\rm d}}
\newcommand{\dd}{{\rm d}}

\def\glu{{\rm g}}
\def\glug{{\rm gg}}
\def\gq{{\rm gq}}
\def\qg{{\rm qg}}
\def\qq{{\rm qq}}

\def\S{{\rm S}}

\newcommand{\nn}{\nonumber\\ }

\def\Raa{R_\mathrm{AA}}

\begin{document}
\title{Multipartonic cascades in expanding media%
\thanks{Presented at Quark Matter 2022, Kraków, Poland}%
}
\author{Souvik Priyam Adhya\thanks{Presenter and corresponding author: {souvik.adhya@ifj.edu.pl}},
\address{Institute of Nuclear Physics, Polish Academy of Sciences,
ul. Radzikowskiego 152, 31-342, Poland}
\\[1mm]
{Carlos A. Salgado
\address{Instituto Galego de F\'isica de Altas Enerx\'ias IGFAE, Universidade de Santiago de Compostela, E-15782 Galicia-Spain}
\\[1mm]
}
{Martin Spousta 
\address{Institute of Particle and Nuclear Physics, Faculty of Mathematics and Physics, Charles University, V Hole\v sovi\v ck\'ach 2, 180 00, Czech Republic}
}
\\[1mm]
Konrad Tywoniuk
\address{Department of Physics and Technology, University of Bergen, 5007, Norway}
}
\maketitle
\begin{abstract}
In this work, we introduce both gluon and quark degrees of freedom for describing the partonic cascades inside the medium. We present numerical solutions for the set of coupled evolution equations with splitting kernels calculated for the static, exponential and Bjorken expanding media to arrive at medium-modified parton spectra for quark and gluon initiated jets respectively. We discuss novel scaling features of the partonic spectra between different types of media. Next, we study the inclusive jet $R_{AA}$ by including phenomenologically driven combinations of quark and gluon fractions inside a jet. In addition, we have also studied the effect of the nPDF as well as vacuum like emissions on the jet $R_{AA}$. Differences among the estimated values of quenching parameter for different types of medium expansions are noted. Next, the impact of the expansion of the medium on the rapidity dependence of the jet $R_{AA}$ as well as jet $v_2$ are studied in detail. Finally, we present qualitative results comparing the sensitivity of the time for the onset of the quenching for the Bjorken profile on these observables. All the quantities calculated are compared with the recent ATLAS data.
\end{abstract}
\section{Theoretical background}
The study of hard probes such as jets and its quenching features inside the hot and dense QCD medium formed in heavy ion collisions such as the LHC and the RHIC can reveal interesting properties of the evolving medium.
The effects of energy loss (quenching of jets) are studied using rate equations describing parton splittings in the medium. The rate itself is derived within the approximation of multiple-soft scattering in expanding media \cite{Adhya:2019qse}, and the in-medium distributions are found using numerical solution of the coupled evolution equations \cite{Mehtar-Tani:2018zba}\cite{Adhya:2021kws}.
We aim is to study the impact of the medium expansion on the jet quenching observables. To compare to  experimental data on jet production in heavy-ion collisions, we have included both gluon and quark degrees of freedom in the description of partonic cascades and emphasize the importance of precise modeling of the input parton spectra.
In comparison to a recent work on gluonic cascades \cite{Adhya:2019qse}\cite{ Adhya:2020vca}, we have included both the gluon and quark degrees of 
freedom for partonic cascades into the evolution equations for a more realistic description of medium evolved parton spectra for an expanding medium \cite{Adhya:2021kws}. Therefore, we are able to provide more precise 
 study of the $\Raa$ and estimate of the quenching parameter, $\hat{q}$, to be extracted from the data. 
  We shall use the optimized values of the jet quenching parameter which minimizes the differences in the inclusive $\Raa$ among different medium profiles to study the inclusive jet suppression in more differential 
way. Namely, we evaluate the rapidity dependence of the $\Raa$ where experiment hints at a non-trivial behavior and then we evaluate the jet $v_2$ constructed from the 
 path-length dependent $\Raa$. This will allow us to quantify the impact of the way how the medium expands on these more differential observable which are being 
 measured by experiments.

In order to describe the in - medium parton cascades, we start from the coupled evolution equations for the in-medium inclusive fragmentation function \cite{Mehtar-Tani:2018zba}\cite{Adhya:2021kws}\cite{Blanco:2021usa},
\begin{align}
\label{eq:evol-eq-glu}
\frac{\del }{\del \tau } D_\glu\left(x,\tau\right)&=\int_{0}^{1} dz \, \Kc_\glug(z) \left[ \sqrt{\frac{z}{x}} D_\glu\left(\frac{x}{z},\tau \right) \Theta(z-x) - \frac{z}{\sqrt{x}} D_\glu(x,\tau) \right] \nn
&-   \int_{0}^{1} \rmd z \,\Kc_\qg(z)  \frac{z}{\sqrt{x}}\, D_\glu\left(x ,\tau \right) +\int_0^1 \rmd z \, \Kc_\gq(z)  \,  \sqrt{\frac{z}{x}} \,D_\S \left(\frac{x}{z},\tau \right) \,,\\
\label{eq:evol-eq-S}
\frac{\del }{\del \tau } D_\S\left(x,\tau\right)&= \int_{0}^{1} dz \, \Kc_\qq(z) \left[ \sqrt{\frac{z}{x}} D_\S\left(\frac{x}{z}, \tau \right) \Theta(z-x) - \frac{1}{\sqrt{x}} D_\S(x,\tau) \right] \nn
& + \int_{0}^{1} dz \, \Kc_\qg(z) \, \sqrt{\frac{z}{x}} D_\glu\left(\frac{x}{z} ,\tau \right) \,.
\end{align}
where $D_g(x)$ and $D_S(x) \equiv \sum_f \big[D_{q_f}(x) + D_{\bar q_f}(x) \big]$ are the gluon and quark singlet distributions respectively. We define $\tau = \sqrt{\hat q_0/E} \, t$ as the evolution variable, $\hat q_0$ as jet quenching coefficient in static medium and $t$ as the distance traversed in the medium. For our medium modelling, we consider three scenarios as follows,
\begin{itemize}
\item{Static medium:} For a static medium of finite medium length $L$, we have $\hat q(t) = \hat q_0 \Theta(L-t)$. The splitting rate is defined as \cite{Arnold:2008iy}\cite{Adhya:2019qse}\cite{Adhya:2021kws},
\beq
\label{eq:rate-static}
\mathcal{K}_{ij}(z,\tau)  =  \frac{\alpha_s}{2\pi} P_{ij}(z) \kappa_{ij} (z)\, \text{Re} \left[(i-1) \tan \left(\frac{1-i}{2} \kappa_{ij} (z) \tau \right) \right]\,.
\end{eqnarray}
In the above equations, $P_{ij}(z)$ is the (unregularised) Altarelli-Parisi splitting functions and $\kappa_{ij}(z)$ is enlisted in Appendix A of \cite{Adhya:2021kws}. We shall also use the case of soft gluon emissions, i.e. $1-z,z \ll 1$ for the static medium.

\item{Exponential expanding medium:} For exponentially expanding media the splitting rate can be derived as \cite{Arnold:2008iy}\cite{Adhya:2019qse}\cite{Adhya:2021kws},
\beq
\label{eq:rate-expo}
\mathcal{K}_{ij}(z,\tau)  = \frac{\alpha_s}{\pi} P_{ij}(z) \kappa_{ij}(z) \,\text{Re} \left[ (i-1) \frac{J_1\big((1-i)\kappa_{ij}(z) \tau \big)}{J_0\big((1-i)\kappa_{ij}(z) \tau \big)} \right] \,.
\eeq
where $\hat q(t) = \hat q_0 e^{-t/L}$ the jet quenching co-efficient.
\item{Bjorken expanding medium:} The jet quenching parameter for the Bjorken expanding medium is defined as,
\beq
\hat q(t) = \begin{cases} 0 & {\rm for } \quad t<t_0 \,, \\ \hat q_0 (t_0/t) & {\rm for} \quad t_0 < t < L+t_0 \,, \\ 0 & {\rm for} \quad L+t_0 < t \,,\end{cases}
\eeq
The splitting rate is written as \cite{Arnold:2008iy}\cite{Adhya:2019qse}\cite{Adhya:2021kws},
\begin{align}
\label{eq:rate-bjorken}
\mathcal{K}_{ij}(z,\tau,\tau_0) &= \frac{\alpha_s}{2\pi} P_{ij} (z) \kappa_{ij} (z)\sqrt{\frac{\tau_0}{\tau + \tau_0}} \nn
&\times \text{Re} \left[ (1-i) \frac{J_1(z_L) Y_1(z_0) - J_1(z_0) Y_1(z_L) }{J_1(z_0) Y_0(z_L) - J_0(z_L) Y_1(z_0)} \right],
\end{align}
where
\begin{align}
z_0 &= (1-i) \kappa_{ij} (z) \tau_0 \,, \\
z_L &= (1-i) \kappa_{ij} (z) \sqrt{\tau_0 (\tau + \tau_0)} \,,
\end{align}
with $\tau_0 = \sqrt{\hat q_0/E} t_0$.
\end{itemize}
 Next, we shall use these medium modified parton splitting functions for all possible combinations of parton splittings to numerically solve the medium evolved parton spectra \cite{Adhya:2021kws} to arrive at quenching factor of the jets in medium.

Thus, the quenching factor can be written as,
\beq
\label{eq:suppression-factor-2}
{\cal Q}_{\{i= q/g\}}(\pT) = \int_0^1 \dd x \,x^{n_i(p_T,y)-1} D(x, \sqrt{x} \tau;\{i\}) \,.
\eeq
Subsequently, combining the contributions from the quark and gluons, the nuclear modification factor $\Raa$ is written as \cite{Adhya:2021kws}, 
\begin{eqnarray}
\label{eq:combRaa}
\Raa &=& \frac{\sigma_q^{\rm 0,med}(\pT,y,R)}{\sigma_q^{\rm 0}(\pT,y,R) + \sigma_g^{\rm 0}(\pT,y,R)}\,{\cal Q}_q(\pT,R) \nonumber\\ &+&\frac{\sigma_g^{\rm 0,med}(\pT,y,R)}{\sigma_q^{\rm 0}(\pT,y,R) + \sigma_g^{\rm 0}(\pT,y,R)} \, {\cal Q}_g(\pT,R) \,,
\end{eqnarray}
where $\sigma_i^{\rm 0,med}$ is vacuum cross-section with nPDF effects and jet cone-size $R$.

Finally, the path-length dependence of the jet quenching phenomena can be explored through the jet $v_2$ as \cite{Zigic:2018smz},
\begin{equation}
\label{eq:v2}
v_2 = \frac{1}{2} \frac{\Raa(L^{in}) - \Raa(L^{out})}{\Raa(L^{in}) + \Raa(L^{out})},
\end{equation}
 where $L$ are the medium lengths traversed by the jet particle in the direction along $(in)$ or perpendicular $(out)$ to the event plane. The values of 
$L^{in}$ and $L^{out}$ are listed in ref.\cite{Adhya:2021kws}.
\begin{figure}
\centering
  \includegraphics[width=60mm]{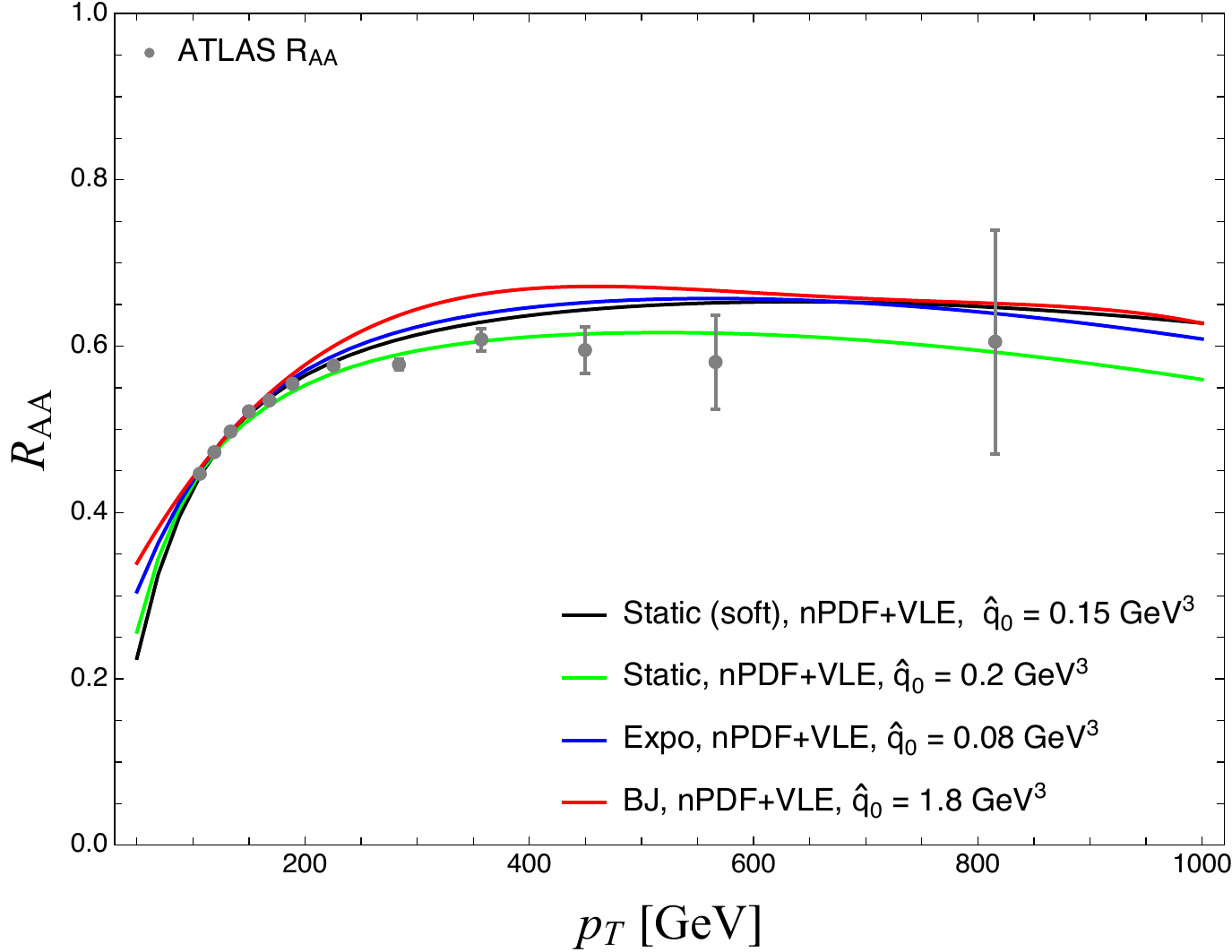}~~~~~\includegraphics[width=40mm]{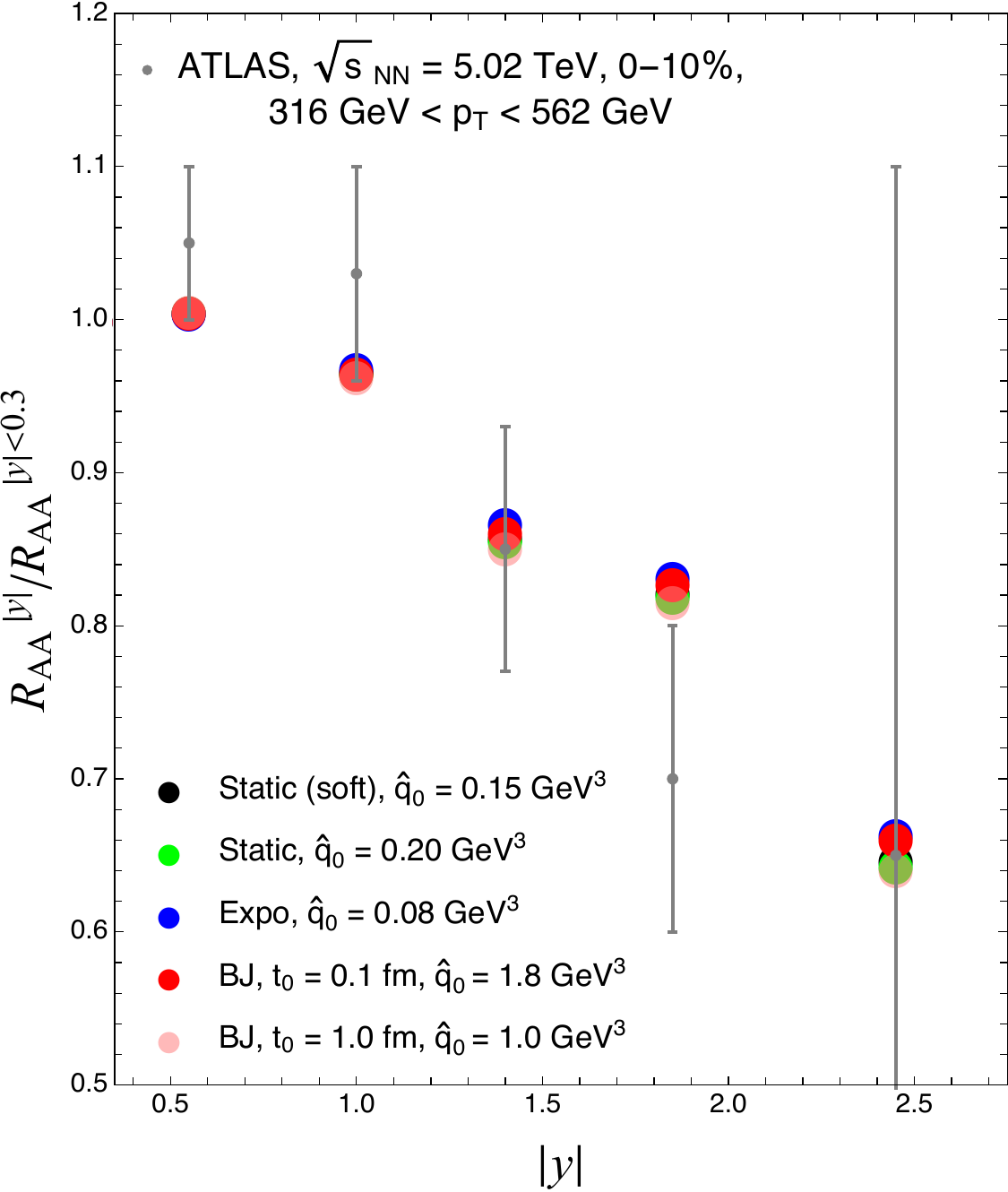}
\caption{(Left panel) The comparison of the jet $\Raa$ for four medium profiles with the ATLAS data (in gray)\cite{Aaboud:2018twu}. (Right panel) The rapidity ratio of $\Raa$ in different $|y|$ bins and $\Raa$ in $|y|<0.3$ for $\pt = 316-512$~GeV. 
 }
\label{fig:Qratiorapsopti}
\end{figure}

\begin{figure}
\centering
	\includegraphics[width=60mm]{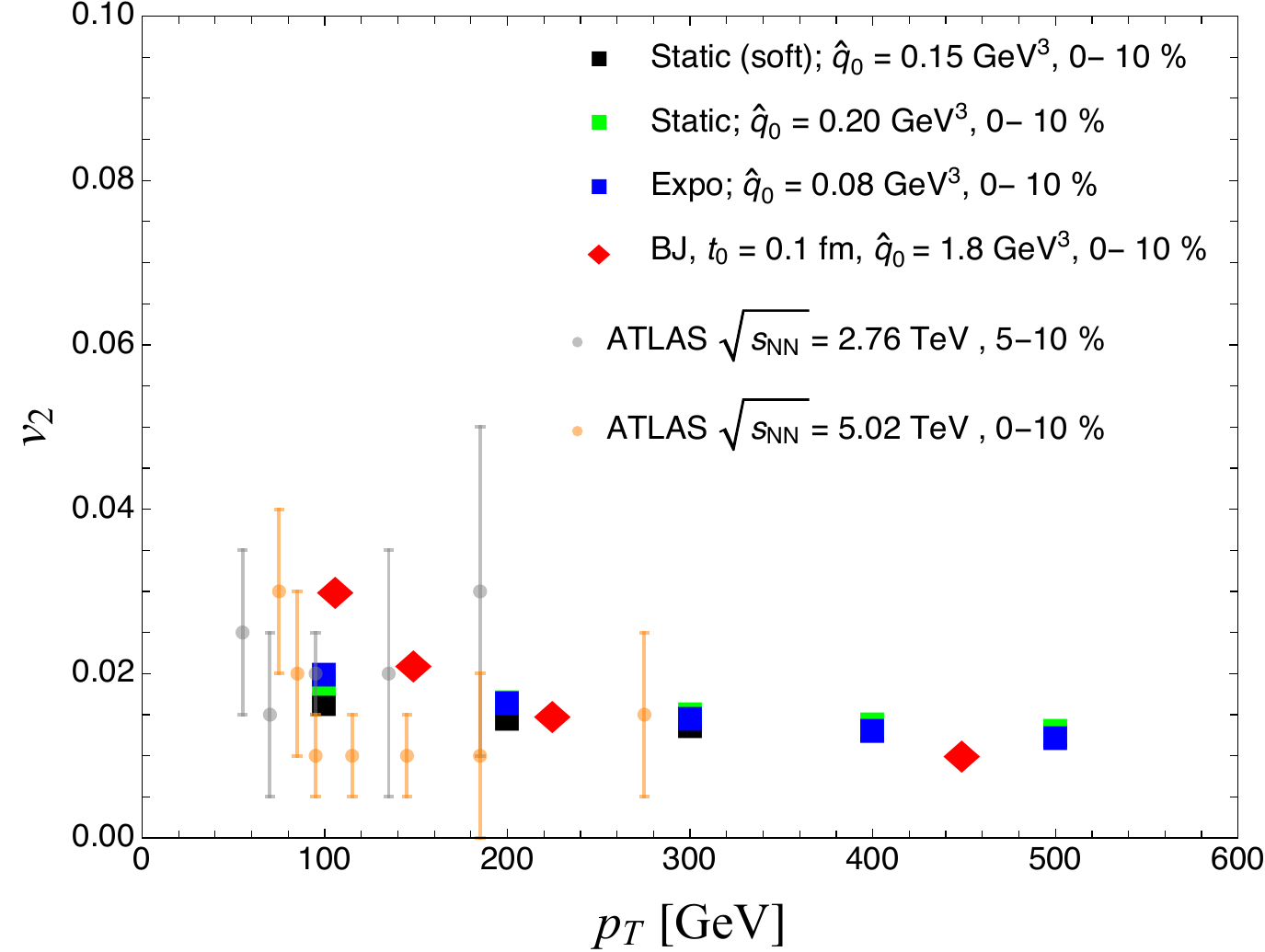}~~\includegraphics[width=60mm]{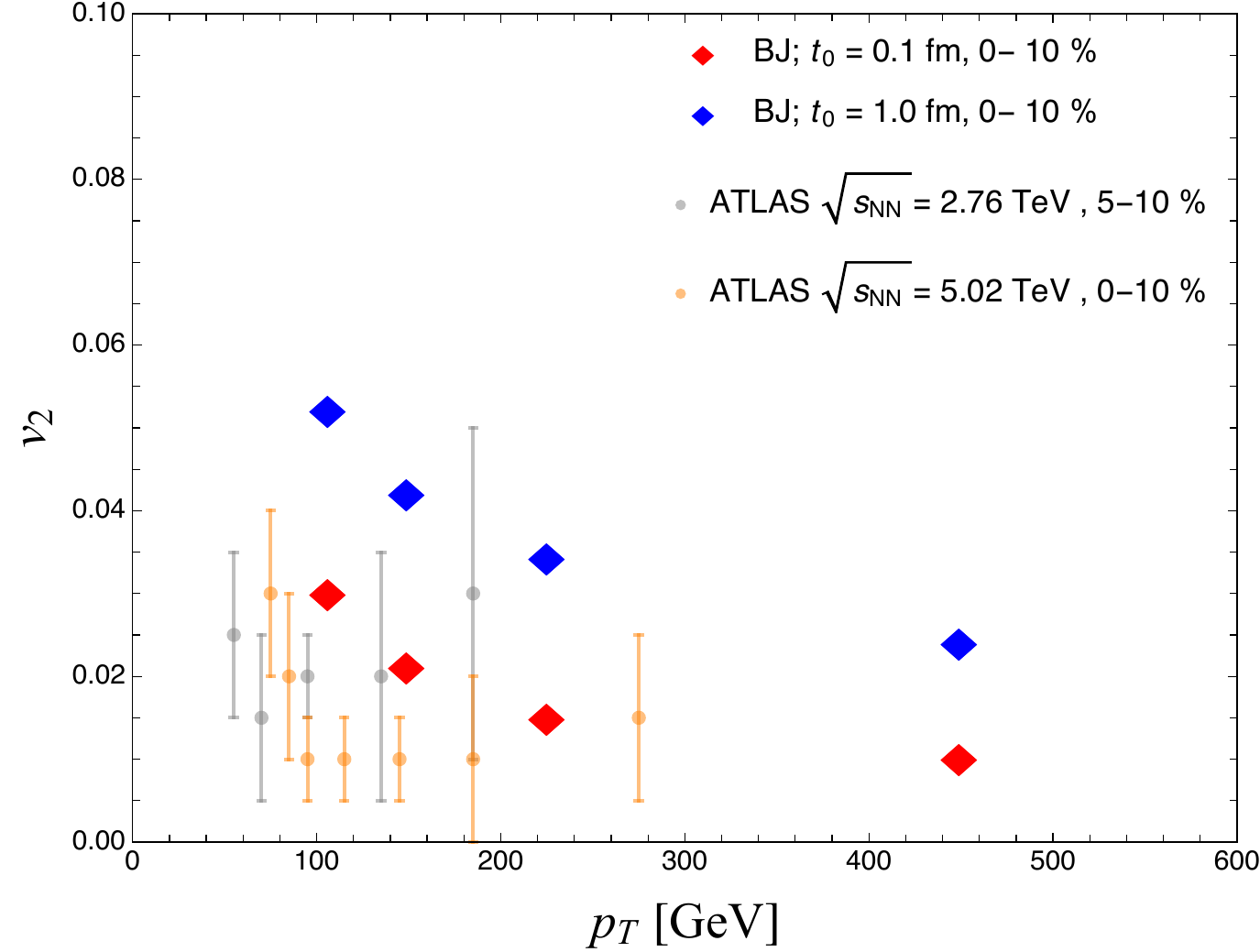}
\caption{
Comparison of the jet $v_2$ for static medium, exponential medium, and Bjorken medium with $t_0 = 0.1$~fm (left panel). Similar comparison (right panel) for Bjorken media for 
$t_0=0.1$~fm (early quenching) and $t_0=1.0$~fm (late quenching). ATLAS data from \cite{ATLAS:2013ssy}\cite{ATLAS:2020qxc}.
}
\label{fig:v2allprofs}
\end{figure}
\section{Results and conclusions}
We present the comparisons among different medium profiles for the $\Raa$ as a function of $p_T$ and $\Raa$ ratio as a function of rapidity in fig.\ref{fig:Qratiorapsopti}. We have used the optimised values of $\hat q_0 $ as outlined in tab. 2 of \cite{Adhya:2021kws}. 
For our purpose, we have included both the nPDF as well as vacuum like emissions (VLE) effects for the input parton spectra for all the medium profiles \cite{Adhya:2021kws}.
In fig. \ref{fig:Qratiorapsopti} (left), although the differences in the magnitude and shape of $\Raa$ among different medium profiles are small, they differ considerably in the optimised $\hat q_0$ values. Inclusion of the quark along with the gluon jets improves the level of completeness of the jet energy loss prescription for all the medium profiles. The extracted  $\hat q_0$ values differs with the ones in \cite{Adhya:2019qse} leading to breakdown of the effective scaling laws observed in \cite{Adhya:2019qse}.
Next, in fig.\ref{fig:Qratiorapsopti} (right), we plot the $\Raa$ ratio for different medium profiles as a function of rapidity and compare with the ATLAS data \cite{Aaboud:2018twu}. Thus, we see that the rapidity dependence does not allow to distinguish between static, exponential, and Bjorken medium profiles \cite{Adhya:2021kws}. The trends seen in the ratio of $\Raa$ in rapidity are the same for all the medium profiles. We conclude that the rapidity dependence is due to change in the steepness of input parton spectra \cite{Adhya:2021kws}. Next, in fig.\ref{fig:v2allprofs}, we study the jet $v_2$ with $p_T$ for different medium profiles compared with measurements from refs.\cite{ATLAS:2013ssy}\cite{ATLAS:2020qxc}. We note a difference of the Bjorken profile with $t_0=0.1$ $fm$  with the static and exponential profiles. A factor of two difference can also be observed between the two Bjorken profiles implying that the jet $v_2$ is sensitive to choice of $t_0$ \cite{Adhya:2021kws}\cite{Andres:2019eus}.
These results of jet rapidity dependence and  $v_2$ presented highlights the sensitivity of the medium expansion on the observed phenomena of jet quenching in heavy ion collisions. 
\section*{Acknowledgements}
KT acknowledges BFS2018REK01 and the University of Bergen. CAS acknowledges EU Horizon 2020 (82409); ERDF; MDM-2016-0692, FPA2017-83814-P and ERC-2018-ADG-835105 YoctoLHC. MS acknowledges GACR 22-11846S and UNCE/SCI/013. SPA acknowledges the European Union’s Horizon 2020 research and innovation program under the Marie Sklodowska- Curie grant agreement No. 847639 and from the Ministry of Education and Science (PAN.BFB.S.BDN.612.022.2021, 2021
- PASIFIC 1, QGPAnatomy).

\bibliographystyle{elsarticle-num}
\bibliography{multpart_expmed} 
\end{document}